\documentclass{article}
\usepackage{spconf,amsmath,graphicx}
\usepackage{booktabs}
\usepackage{makecell}

\usepackage{fancyhdr}

\title{AUTOSEN: IMPROVING AUTOMATIC WIFI HUMAN SENSING THROUGH CROSS-MODAL AUTOENCODER}
\name{Qian Gao, Yanling Hao, Yuanwei Liu}
\address{Queen Mary University of London, London, UK}

\fancypagestyle{copyrightpage}{
	\fancyhf{}  
	\fancyfoot[L]{© 20XX IEEE. Personal use of this material is permitted. Permission from IEEE must be obtained for all other uses, in any current or future media, including reprinting/republishing this material for advertising or promotional purposes, creating new collective works, for resale or redistribution to servers or lists, or reuse of any copyrighted component of this work in other works.}
}

\begin{document}
%
\maketitle
\begin{abstract}

\thispagestyle{copyrightpage}
	
WiFi human sensing is highly regarded for its low-cost and privacy advantages in recognizing human activities. However, its effectiveness is largely confined to controlled, single-user, line-of-sight settings, limited by data collection complexities and the scarcity of labeled datasets. Traditional cross-modal methods, aimed at mitigating these limitations by enabling self-supervised learning without labeled data, struggle to extract meaningful features from amplitude-phase combinations. In response, we introduce AutoSen, an innovative automatic WiFi sensing solution that departs from conventional approaches. AutoSen establishes a direct link between amplitude and phase through automated cross-modal autoencoder learning. This autoencoder efficiently extracts valuable features from unlabeled CSI data, encompassing amplitude and phase information while eliminating their respective unique noises. These features are then leveraged for specific tasks using few-shot learning techniques. AutoSen's performance is rigorously evaluated on a publicly accessible benchmark dataset, demonstrating its exceptional capabilities in automatic WiFi sensing through the extraction of comprehensive cross-modal features.

\end{abstract}
\section{Introduction}

WiFi human sensing uses channel state information (CSI) to measure the changes in reflection, diffraction, and scattering paths caused by human activities. Based on IEEE 802.11 n/ac \cite{csisurvey} standards, CSI is a matrix of complex variables with spatial diversity due to multiple antennas in Multiple-Input Multiple-Output (MIMO) setup and frequency diversity due to multiple subcarriers in Orthogonal Frequency Division Multiplex (OFDM) modulation. Different actions will cause CSI to be distinguished by various changes in antennas and subcarriers.

The solutions for human activity recognition can be categorized as model-based and pattern-based approaches \cite{modelsurvey}. The model-based approach tries to find a physical model to accurately describe the electromagnetic wave propagation during human motion. Although this method has achieved extremely high accuracy on some specific tasks, like respiration detection \cite{respiration} and walk direction estimation \cite{walk}, it is only suitable for periodic repetitive motions and has very strict requirements on the placement of devices. While the pattern-based approach can be used for more complicated tasks by utilizing advanced artificial intelligence technology. Compared to machine learning methods, such as random forest (RF), support vector machine (SVM) \cite{usenexmon1}, and hidden Markov model (HMM) \cite{csisurvey}, deep learning models \cite{twostream} provide better feature extraction and end-to-end model learning capability.  

Although WiFi human sensing has made great progress in recent years. It's too labor-intensive to collect annotated data in new scenarios. To overcome this shortage, in AutoFi \cite{autofi}, authors propose a framework to learn CSI features from unlabeled data using geometric self-supervised learning \cite{byol} and then transfer the knowledge to specific tasks defined by users through few-shot learning \cite{fewshot}. 

\begin{figure*}[htbp]  
	\centering
	\includegraphics[width=0.95\linewidth,scale=1.00]{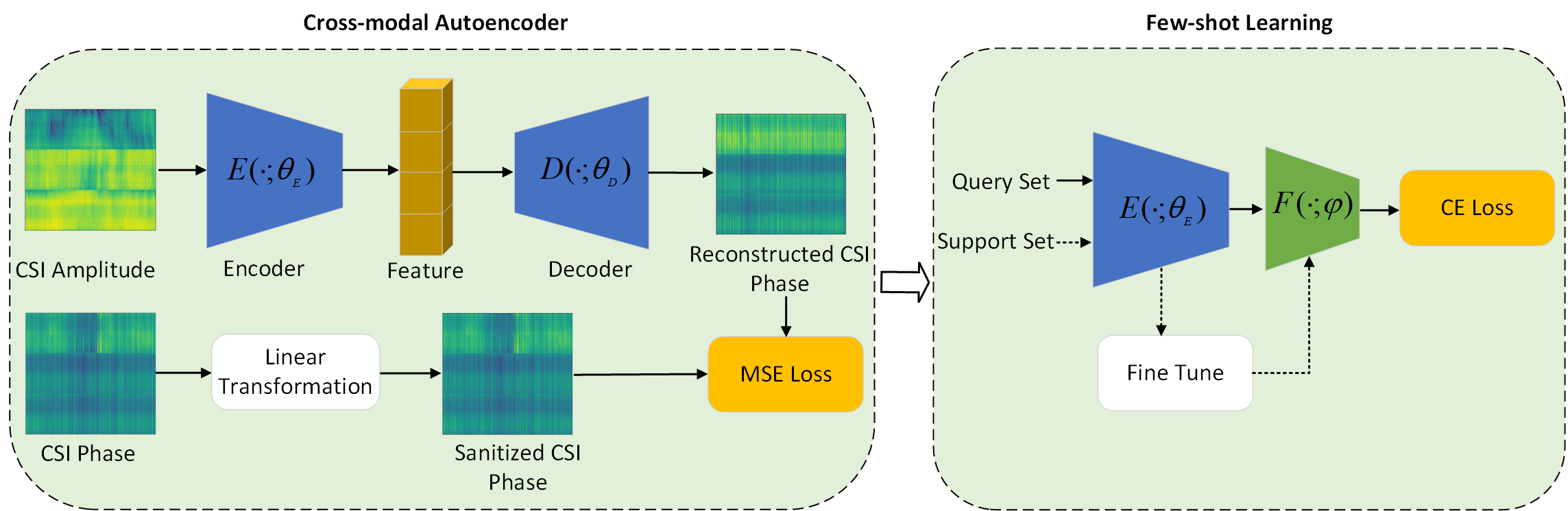}
	\caption{Overview of proposed AutoSen model. The AutoSen consists of a cross-modal autoencoder module that extracts features from unlabeled CSI amplitude and phase, and a few-shot learning module that transfers the knowledge to specific tasks. Note that the encoder in few-shot learning is the one obtained in cross-modal autoencoder. }
	\label{fig1}	
\end{figure*}  

AutoFi \cite{autofi} gives a first step towards automatic WiFi sensing. However its performance is restricted by the limited information in CSI amplitude. CSI phase is another representation modality \cite{mo} that is orthogonal to corresponding amplitude and contributes to human activity recognition. While raw CSI phase contains undesired offsets, like channel frequency offset (CFO), sampling frequency offset (SFO), and packet detection delay (PDD), there have been proposed many sanitization methods \cite{linear, antenna, static, sharp} to obtain cleaned CSI phase.

In this paper, we propose AutoSen to improve automatic WiFi human sensing from the perspective of feature engineering. A convolutional autoencoder \cite{efficientfi} is employed to extract cross-modal representations from CSI amplitude and sanitized phase. With the trained encoder, the recognition performance of downstream tasks based on few-shot learning is boosted closer to its upper boundary. The main contributions are summarized as follows:

1) We propose an automatic WiFi human sensing model (AutoSen) to train more effective encoder without requirement of annotated data.

2) We design a cross-modal autoencoder module to directly bridge the CSI amplitude and phase, which describes motions from different aspects while eliminating unique noises in these two modalities.

3) Experiments on a public dataset show the effectiveness of cross-modal features and our proposed AutoSen outperforms other single-modal unsupervised methods on recognition accuracy without increasing the model size.

\section{AutoSen  Model}

 Our proposed AutoSen consists of two parts as seen in Fig.\ref{fig1}: a cross-modal autoencoder module and a few-shot learning module. The details are described as follows.

\subsection{WiFi Sensing Basics}

CSI amplitude describes human activities from the perspective of the extent that the human body reflects specific multipath signals, while corresponding CSI phase reflects the delays caused by the length change of paths. But both of these two properties contain the features relevant to human motions. Fig.\ref{fig2} illustrates how CSI changes with human activity. It can be observed that as a person moves from the blue figure to the orange figure, the propagation path length increases, and the signal changes from being reflected by the human body to being reflected by the ground (The orange solid line and the blue dashed line emanate from the transmitter at the same angle, $\theta = \theta^{\prime}$). Consequently, both the magnitude and phase of the CSI undergo changes. For commercial off-the-shelf WiFi NIC, the CSI can be modeled as:

\begin{figure}[htbp]  
	\centering
	\includegraphics[width=0.8\linewidth,scale=1.00]{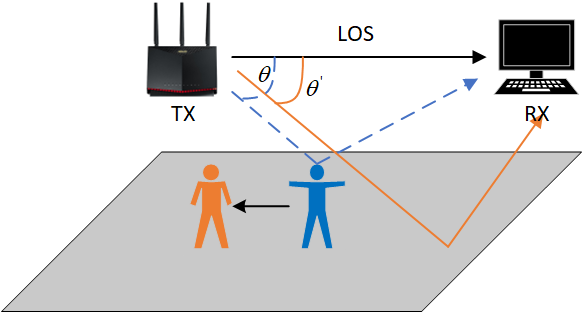}
	\caption{Illustration of how CSI changes in response to human activity. }
	\label{fig2}	
\end{figure}  

\begin{equation}
	\label{1}
	\begin{aligned}
		H_{i}(n)&=A_{i}(n)e^{j\phi_{i}(n)} \\
		&=\sum_{p=0}^{P-1}A_{p}(n)e^{j2\pi(f_{c}+k/T)\tau_{p}(n)},
	\end{aligned}
\end{equation}
where $i \in \{1, \dots, K\}$, $n \in \{1, \dots, N\}$, and $p \in \{1, \dots, P\}$ are subcarrier, packet, and multipath number respectively, with $f_{c}$ being the main carrier frequency, while $T=1/\Delta f$ is the OFDM symbol time and $\tau_{p}(n)$ is the time delay of $n$th packet on $p$th subcarrier.

Our insight is that CSI amplitude and phase can be viewed as two kinds of modalities \cite{mo} and we are able to obtain superior cross-modal representatives by utilizing an autoencoder.

\begin{table*}[htbp]
	\centering
	\small
	\caption{The network architecture used in the AutoSen experiments. For Conv $A×(H,W)$, A denotes the channel number, and $(H,W)$ represents the height and width of the operation kernel. This applies to all Convolution (Conv) and ConvTranspose (ConvTrans) layers.
	}
	\begin{tabular}{c|c|c|c}
		\hline
		Layer Index & Encoder $E$ & Decoder $D$ & Classifier $F$ \\
		\hline
		input & \multicolumn{3}{c}{CSI data: $1 \times 500\times 90$ ($antenna \times timestamp \times subcarrier$)} \\
		\hline
		1     & Conv $32\times(10,5)$, stride (10,5), ReLU &$96\times 4\times6 $ dense & 256 dense \\
		\hline
		2     & Conv $64\times(10,3)$, stride (10,3), ReLU & ConvTrans $96\times(5,1)$, stride (5,1), ReLU & 128 dense \\
		\hline
		3     & Conv $96\times(5,1)$, stride (5,1), ReLU & ConvTrans $64\times(10,3)$, stride (10,3), ReLU & 7 dense, softmax\\
		\hline
		4     & 256 dense & ConvTrans $32\times(10,5)$, stride (10,5), ReLU &\\
		\hline
	\end{tabular}%
	\label{1}%
\end{table*}%

\subsection{Cross-modal Autoencoder Module}

In our model, we extract cross-modal features via a convolution autoencoder. To be  specific, CSI amplitude is first fed into an encoder $E(\cdot;\theta_{E})$ to obtain the latent feature representation $h$ that illustrates the information about human activities:
\begin{equation}
	h=E(CSI_{amp}; \theta_{E}),
\end{equation}
Then the decoder $D(\cdot;\theta_{D})$ uses $h$ to reconstruct CSI phase rather than its amplitude: 
\begin{equation}
	\widetilde{CSI_{pha}}=D(h; \theta_{D}),
\end{equation}
Note that raw CSI phase has unwanted offsets, we sanitize the raw CSI phase with a linear transformation mentioned in \cite{linear}:

\begin{equation}
	CSI_{pha_i}=\widehat{CSI_{pha_i}}-km_{i}-b,
\end{equation}
where $k$ and $b$ respectively denote the slope and offset of phase across the entire frequency band, and $m_{i}$ is the subcarrier index of $i$th subcarrier. 
Next, the goal of cross-modal autoencoder module is to minimize the Mean Square Error (MSE) between reconstructed CSI phase and sanitized CSI phase:

\begin{equation}
	L_{MSE}=\frac{1}{M}\sum_{i=1}^{M}\left| \widetilde{CSI_{pha}}-CSI_{pha}  \right|^{2},
\end{equation}
Here, $M$ is the number of unlabeled CSI amplitude and phase pairs.

\subsection{Few-shot Learning Module}

With the cross-modal autoencoder module, the encoder $E(\cdot;\theta_{E})$ learns how to extract activity relevant features from CSI amplitude directly.  To transfer the knowledge to downstream tasks, we employ a few-shot learning module, in which we use a few labeled samples $(x,y)$ and the same encoder $E(\cdot;\theta_{E})$ to train a classifier $F(\cdot;\varphi)$ for user-defined tasks by minimizing the Cross Entropy (CE) loss:

\begin{equation}
	L_{CE}=-\mathbf{E}_{\left(x,y\right)}\sum_{c}\left[ \mathbf{I}\left[ y=c\right] log \left( F(E(x;\theta_{E});\varphi) \right) \right],
\end{equation}
where $\mathbf{E}$ represents expectation and $\mathbf{I}$ is a 0-1 function that outputs 1 for correct label $c$. Finally, the actual label of a human activity can be predicted by:
\begin{equation}
	label=argmax\left( F(E(x;\theta_{E});\varphi) \right).
\end{equation}

\section{Experiments}

\subsection{Data}

\begin{table}[htbp]
	\centering
	\small
	\caption{Accuracy (\%) comparison on UT\_HAR \cite{csisurvey} dataset. FS is short for Full Supervision, amp and pha are short for amplitude and phase, respectively.}
	\begin{tabular}{c|c|c|c|c}
		\hline
		\multicolumn{2}{c|}{Method} & 10-shots & 20-shots & Avg \\
		\hline
		supervised& \makecell[c]{FS (amp) \\FS (pha)}&   \makecell[c]{92.86\\53.24}    &  \makecell[c]{94.23 \\53.58}    & \makecell[c]{93.55\\53.41} \\
		
		\hline
		
		unsupervised& \makecell[c]{AutoFi (amp) \\AutoFi (pha)\\\textbf{AutoSen}}&   \makecell[c]{67.63\\32.37\\\textbf{  71.32 }}    &  \makecell[c]{75.16 \\34.13\\\textbf{ 78.92 } }    & \makecell[c]{71.40\\ 33.25\\\textbf{75.12}} \\
		
		\hline
	\end{tabular}%
	\label{2}%
\end{table}%

In this paper, we used the UT\_HAR dataset provided in \cite{csisurvey}. The transceivers with Intel 5300 NIC are placed 3 meters away in an office. There are three antennas at receiver and each antenna has 30 subcarriers. Six persons perform seven activities, i.e., bed, fall, pickup, run, sitdown, standup and walk for 20 times at the sample rate of 1k Hz. There are 4973 CSI samples in total. Follow the AutoFi \cite{autofi}, we randomly segmented the continuous dataset to imitate the randomly-collected CSI data and employed them for unsupervised training. Then, 10 and 20 labeled samples each category are used for few-shot calibration and 70 samples each category are used for evaluation.

\subsection{Setup}

The proposed AutoSen model is implemented with pytorch on an Nvidia A40 GPU. The CSI data are down-sampled to 500Hz and there are no more preprocess expect to the phase sanitization. The learning rate is 0.001 and optimizer is Adam. The training epoch are both 100 for cross-modal autoencoder module and few-shot learning module, and the batchsize is 64. The details of encoder, decoder, and classifier are shown in Table \ref{1}.

The purpose of AutoSen is to improve the recognition performance of automatic WiFi sensing proposed in AutoFi \cite{autofi} by employing cross-modal autoencoder. The baseline methods are AutoFi and its supervised learning counterpart, Full Supervision, which replaces the geometric self-supervised learning with a supervision network as the upper bound. For fairness, the feature extractor and classifier used in AutoFi and Full Supervision have the same architecture with the encoder and classifier used in AutoSen.

\begin{table}[htbp]
	\centering
	\small
	\caption{Ablation experiments on the influence of modalities on AutoSen.}
	\begin{tabular}{c|c|c|c|c}
		
		\hline
		Modality & 10-shots & 20-shots & Avg & $\Delta$ \\
		\hline
		Amplitude Only & 64.37 & 72.52 & 68.63 & -6.49 \\
		Sanitized Phase Only & 35.16 & 37.26  & 36.21  & -38.91\\
		\makecell[c]{Amplitude \& Phase \\ (Concatenation)} & 68.42 & 73.53 & 70.98  &  -4.15\\
		\textbf{\makecell[c]{Amplitude \& Phase \\ (Cross-modal)}} & \textbf{71.32} & \textbf{78.92} & \textbf{75.12} & \textbf{0}\\
		\hline
	\end{tabular}
	\label{3}
\end{table}%

\subsection{General Performance Analysis}

In Table \ref{2}, we compared the top-1 accuracy of proposed AutoSen and four baselines, i.e. amplitude and phase versions of Full Supervision and AutoFi models, on 10-shots and 20-shots tasks. The results show that:

1) Full Supervision (amplitude) trains the encoder in a supervised manner and the upper boundaries of UT\_HAR dataset are 92.86\% and 94.23\% for 10-shots and 20-shots tasks, respectively. Our model AutoSen achieves the accuracy of 71.32\% and 78.92\% for two tasks, which are 3.69\% and 3.72\% higher than accuracies of AutoFi (amplitude). This demonstrates the effectiveness AutoSen on HAR tasks. The cross-modal features are more representative than those learned by single modality.

2) When we changed the inputs from amplitude into phase, the performance degrades a lot. The average accuracy of AutoFi (phase) is 33.25\% while it is 53.41\% for Full Supervision. There are two reasons behind this phenomenon. Firstly, phase is more unstable than amplitude and sanitization can not remove all the offsets inside phase. Secondly, the resolution of UT\_HAR is small due to the fact that Intel 5300 NIC \cite{nic} only support 30 subcarriers for each antenna of receiver.

3) For amplitude input methods and proposed AutoSen, the accuracies are improved around 8\% when the shots increased from 10 to 20. While for phase input counterparts, their accuracies are almost the same when the number of shots change. This again verifies the negative impact of offsets in phase and low resolution data on few-shot learning. 

It's important to note that when using CSI extraction tools with a higher number of subcarriers and antennas, our model can achieve an accuracy level exceeding 80\%.

\subsection{Impact of Cross-modal Features}

To investigate the contributions of cross-modal features, we conducted ablation studies on four different modalities. Amplitude only and sanitized phase only mean that we input amplitude (phase) to the encoder and then reconstructed same modality with decoder. Amplitude \& phase (concatenation) is the situation that we directly concatenated CSI amplitude and phase as the input and output of the model. While amplitude \& phase (cross-modal) is our proposed AutoSen, the input amplitude is used to reconstruct its corresponding phase. The results are shown in Table \ref{3}. 

\begin{table}[htbp]
	\centering
	\small
	\caption{Impact of the size of latent feature.}
	\begin{tabular}{c|c|c|c|c}
		\hline
		Latent Feature Size & 10-shots & 20-shots & Avg & $\Delta$ \\
		\hline
		128 &    64.62   &  71.77     & 68.20 & -6.92\\
		
		\textbf{256} &    \textbf{ 71.32}   &  \textbf{ 78.92}    & \textbf{75.12} & \textbf{0} \\
		
		512 &    66.29   &   72.02    &  69.25& -5.87\\
		\hline
	\end{tabular}%
	\label{4}%
\end{table}%

It can be seen that the performance of cross-modal autoencoder has a  6.49\% increase over that of amplitude only method. This is because CSI amplitude and phase describe human motions from different aspects and cross-modal features contains more relative information. Note that the recognition accuracies of cross-modal are not influenced by phase offsets and low resolution mentioned above and have an improvement of 38.91\% on average, while the result of amplitude \& phase (concatenation) is 70.98\%, which has a limited increase compared to amplitude only method. This examines out assumption that amplitude and phase have different kinds of noises and our cross-modal autoencoder realizes a noise reduction between these two modalities.

\subsection{Impact of Latent Feature Size}

The convolution autoencoder will not reconstruct the CSI phase well if the number of neurons in the bottleneck is too small or too large. We tried different latent feature sizes in our experiments to find out the best neuron number. Table \ref{4} illustrates that the best latent feature size is 256 for our cross-modal autoencoder model. The average accuracy of 256 neurons is 75.12\%, which is 6.92\% higher than the accuracy of 128 neurons and 5.87\% higher than the accuracy of 512 neurons. 

This is intuitive because when the bottleneck layer is small, it's hard to extract enough information from input data. While the performance will also decrease due to the CSI data also contains the environment feedback. There is trade-off between the motion-related and environment-related features.

\section{Conclusion}
WiFi sensing holds great promise for cost-effective and privacy-conscious applications in smart homes. While accuracy has been impressive in controlled scenarios, the scarcity of datasets remains a substantial hurdle for real-world deployment. To address this challenge, we proposed AutoSen approach. Unlike conventional domain adaptation methods reliant on annotated data, AutoSen employs a cross-modal autoencoder to extract meaningful features from unlabeled CSI amplitude and phase information, preserving their correlations while mitigating their inherent noises. These features are then applied to specific tasks through few-shot learning. Our evaluation on a public benchmark dataset highlights AutoSen's superior performance in automatic WiFi sensing. By effectively extracting cross-modal features and enabling knowledge transfer, AutoSen offers a promising solution to the dataset shortage problem, thereby enhancing the practicality and generalizability of WiFi sensing technology. 

\vfill\pagebreak

\bibliographystyle{IEEEbib}
\bibliography{refs,strings}

\end{document}